\documentclass[12pt,letterpaper]{article}
\usepackage{amssymb}
\usepackage{amsmath}
\usepackage{amscd}
\usepackage{graphicx}
\usepackage{tikz}
\usepackage{subfigure}
\usepackage{array}
\usepackage{pdflscape}
\usepackage{tensor}
\usepackage{ulem}
\usepackage{slashed}

\def\be{\begin{equation}}
\def\ee{\end{equation}}
\def\bea{\begin{eqnarray}}
\def\eea{\end{eqnarray}}
\def\bse{\begin{subequations}}
\def\ese{\end{subequations}}

\def\bna{\bql\begin{array}{rcl}}
\def\ena{\end{array}\eql}
\def\bnn{\beq\begin{array}{rcl}}
\def\enn{\end{array}\ee}
\def\bet{\begin{tabular}}
\def\bsf{\sffamily\bfseries}

\def\4{\text{\bsf-}}

\definecolor{Red}    {rgb}{1.00,0.00,0.00} %  1  (255,  0,  0)
\definecolor{Green}  {rgb}{0.00,0.75,0.00} %  2  (  0,191,  0)
\definecolor{Blue}   {rgb}{0.00,0.00,1.00} %  3  (  0,  0,255)
\definecolor{Orange} {rgb}{1.00,0.67,0.00} %  4  (255,170,  0)
\definecolor{Purple} {rgb}{0.50,0.00,0.50} %  5  (127,  0,127)
\definecolor{Gold}   {rgb}{1.00,0.90,0.00} %  6  (255,230,  0)
\definecolor{Magenta}{rgb}{1.00,0.00,1.00} %  7  (255,  0,255)
\definecolor{Turque} {rgb}{0.00,0.90,0.90} %  8  (  0,230,230)
\definecolor{Seaweed}{rgb}{0.00,0.25,0.00} %  9  (  0, 63,  0)
\definecolor{Brown}  {rgb}{0.50,0.13,0.00} % 10  (127, 32,  0)
\definecolor{Cobalt} {rgb}{0.00,0.00,0.50} % 11  (  0,  0,127)
\definecolor{Sage}   {rgb}{0.00,0.50,0.38} % 12  (127, 32,  0)
\definecolor{grey1}  {rgb}{0.20,0.20,0.20} % 13
\definecolor{grey2}  {rgb}{0.40,0.40,0.40} % 14
\definecolor{grey3}  {rgb}{0.60,0.60,0.60} % 15
\definecolor{grey4}  {rgb}{0.80,0.80,0.80} % 16
\definecolor{grey5}  {rgb}{0.90,0.90,0.90} % 17
\def\C#1#2{{\ifcase#1\or%Greg's color scheme
             \color{Red}\or\color{Green}\or\color{Blue}\or
              \color{Orange}\or\color{Purple}\or\color{Gold}\or
             \color{Magenta}\or\color{Turque}\or\color{Seaweed}\or
               \color{Brown}\or\color{Cobalt}\or\color{Sage}\or
                 \color{grey1}\or\color{grey2}\or\color{grey3}\or
                 \color{grey4}\else\color{grey5}\fi#2}}
\definecolor{gray}{rgb}{.7,.7,.7}
\def\XXX{\colorbox{yellow}{\color{red}\bf X\kern-4pt{\Large$\bs*$}\kern-4.125ptX}}
\def\[{\left[}
\def\]{\right]}

\font\ro=cmsy10                          % font with rope
\def\kcr{{\hbox{\ro \char'170}}}                % right-handed rope
\def\ktl{{\hbox{\ro \char'170}}}        % top end for left-handed rope
\def\ktr{{\hbox{\ro \char'170}}}        % " right
\def\kbl{{\hbox{\ro \char'170}}}        % " bottom left
\def\kbr{{\hbox{\ro \char'170}}}        % " right
\parskip=\medskipamount                 % space between paragraphs (LaTeX)
\thicklines                         % thick straight lines for pictures (LaTeX)

\newskip\humongous \humongous=0pt plus 1000pt minus 1000pt

\newif\ifdtup

\def\border{                                            % border
        \setlength{\unitlength}{1mm}
        \newcount\xco
        \newcount\yco
        \xco=-21
        \yco=12
        \begin{picture}(140,0)
        \put(\xco,\yco){$\ktl$}
        \advance\yco by-1
        {\loop
        \put(\xco,\yco){$\kcr$}
        \advance\yco by-2
        \ifnum\yco>-240
        \repeat
        \put(\xco,\yco){$\kbl$}}
        \xco=158
        \yco=12
        \put(\xco,\yco){$\ktr$}
        \advance\yco by-1
        {\loop
        \put(\xco,\yco){$\kcr$}
        \advance\yco by-2
        \ifnum\yco>-240
        \repeat
        \put(\xco,\yco){$\kbr$}}
        \put(-19.75,13){\tiny **University of Maryland * Center for String and
         Particle  Theory * Physics Department**University of Maryland * Center
        for String and Particle  Theory** }
        \put(-19.75,-241.5){\tiny **University of Maryland * Center for String and
         Particle  Theory * Physics Department**University of Maryland * Center
        for String and Particle  Theory** }
        \end{picture}
        \par\vskip-8mm}

\def\headpic{                                           % UM heading
        \indent
        \setlength{\unitlength}{.4mm}
        \thinlines
        \par
        \begin{picture}(29,16)
        \put(165,16){\line(1,0){4}}
        \put(170,16){\line(1,0){4}}
        \put(180,16){\line(1,0){4}}
        \put(175,0){\line(1,0){4}}
        \put(180,0){\line(1,0){4}}
        \put(185,0){\line(1,0){4}}
        \put(169,0){\line(0,1){16}}
        \put(170,0){\line(0,1){16}}
        \put(179,0){\line(0,1){16}}
        \put(180,0){\line(0,1){16}}
        \put(184,0){\line(0,1){16}}
        \put(185,0){\line(0,1){16}}
        \put(169,16){\oval(8,32)[bl]}
        \put(170,16){\oval(8,32)[br]}
        \put(179,0){\oval(8,32)[tl]}
        \put(185,0){\oval(8,32)[tr]}
        \end{picture}
        \par\vskip-6.5mm
        \thicklines}
\def\endtitle{\end{quotation}\newpage}                  % end title page

\begin{document}

\border\headpic {\hbox to\hsize{\today \hfill
{PP 012-017}}}
\par \noindent
{ \hfill
{hep-th/xxxx.xxxx}}
\par

\par

\setlength{\oddsidemargin}{0.5in}
\setlength{\evensidemargin}{-0.5in}
\begin{center}
\vglue .10in
{\large\bf The Spectrum Of Hypercubes Quotiented By Doubly Even Codewords And The Thermodynamics Of Adinkras}\\[.8in]

Keith\, Burghardt\footnote{keith@umd.edu}
\\[1.7in]

{\it Center for String and Particle Theory\\
Department of Physics, University of Maryland\\
College Park, MD 20742-4111 USA}\\[1in]

{\bf ABSTRACT
}
\\[.01in]
\end{center}
\begin{quotation}
{
In a previous paper, a solution to the problem of determining isomorphism classes of Lie algebra representations was explored using graphs called adinkras and subgraphs called baobabs \cite{Baobab}. In this paper, I show that adinkras contain Shannon entropy and a latent heat from the information stored in their associated baobabs. In Garden algebra, both properties are closely related to the spectrum of hypercubes quotiented by doubly even codewords, which is introduced in this paper.
}

%${~~~}$ \newline
%PACS: 04.65.+e
\endtitle

%%%%%%%%%%%%%%%%%%%%%%%%%%%%%%%%%%%%%%%%%%%%%%%

\begin{center}
${~~~}$ \newline
``$\text{\it{\small Make things as simple as possible, but not simpler.}}$"${~~~}$
\end{center}
$~~~~~~~~~~~~~~~~~~~~~~$ {\small -- Albert Einstein}
\newline ${~~~}$
\noindent

\section{Introduction}
$~~\ ~$  Adinkras are graphs with distinct edge colors and edge weights that encode infinitesimal generator representations \cite{YanThesis} \cite{Adinkra} \cite{Baobab}. Special subgraphs that contain all of the degrees of freedom of an adinkra are called baobabs. Baobabs are used to differentiate isomorphism classes of representations of Super Lie algebras via their encoded degrees of freedom\footnote{Previous work has shown that determining the isomorphism class of an adinkra $\implies$ finding the isomorphism class of the associated baobab \cite{Baobab}.}. Many baobabs can create a single adinkra, although if baobabs with the same topology and edge colors (i.e. {\it chromotopology}) are observed among two adinkras, then their isomorphism class can be rapidly determined \cite{Baobab}.

In this paper, we will focus on a type of adinkra that is well known for classifying supersymmetry in 1 dimension, called a Garden adinkra\footnote{To better understand Garden adinkras,  see\cite{YanThesis} or \cite{Adinkra}. The former characterizes these graphs from a mathematician's perspective, while the latter, from the perspective of physicists.}.
In \cite{Baobab}, it was found that traversing any spanning tree and a few well-chosen cycles of a Garden adinkra will determine all of its properties. In this paper, the author will determine the surjective nature of baobabs in order to better understand the isomorphism class of baobabs themselves, as well as the energy needed to create an adinkra from a baobab. This energy can be considered the latent heat of the phase transition between baobabs and adinkras, introducing a previously unexplored thermodynamic interpretation of adinkras.

It is rather non-obvious to determine the number of baobabs in a given adinkra, because one must know the spectrum first. The topology of a Garden adinkra, however, is that of a quotiented hypercube, therefore, the well known spectrum of a hypercube has been used in part to determine this new spectrum. 

The paper will be organized as follows: section 2 will give an overview of baobabs and adinkras in Garden algebra. Section 3 will determine the spectrum of hypercubes quotiented by doubly even codewords, which are used to bound the multiplicity of baobabs, ignoring directed edges. Section 4 will introduce some thermodynamic properties of adinkras, and section 5 will conclude the paper. Furthermore, an appendix is provided with the proof of the doubly-even quotiented hypercube spectrum.  For the rest of the paper, a ``quotiented hypercube" is assumed to be quotiented by doubly even codewords.

\section{Preliminaries}
$~~\ ~$ The elements of Garden algebra can be represented as a graph (a {\it Garden  adinkra}), such as Fig. \# 1, where the adjacency matrix of a distinct edge color (or {\it adinkra adjacency matrix}) corresponds to an element's representation, with the appropriate non-zero matrix elements. 
\begin{figure}[!htbp]
\centering
\subfigure[]{\label{f:diamondPath}
\includegraphics[width=0.3\columnwidth]{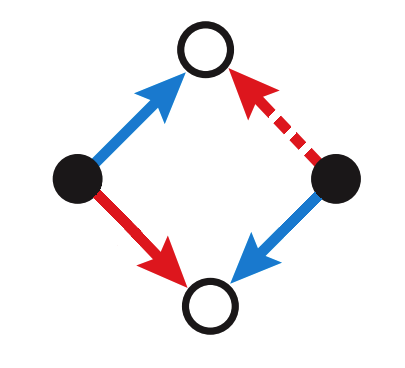}}
\put(-47,3){$2$}      \put(-47,97){$1$}
\put(-106,59){$1$}      \put(-10,59){$2$}
\quad
\subfigure[]{\label{f:bowtiePath}
\includegraphics[width=0.3\columnwidth]{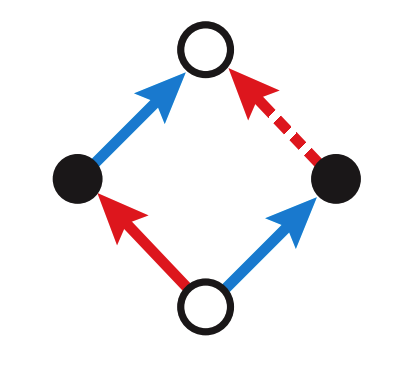}} 
\put(-47,3){$2$}         \put(-47,97){$1$}
\put(-106,59){$1$}      \put(-10,59){$2$}
\label{f:diambtPath}
\caption{Two-color adinkras with labels for bosons (open nodes) and fermions (closed nodes).}
\end{figure}
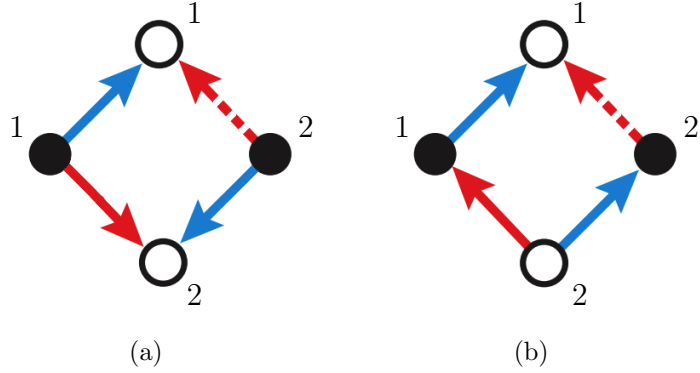

A garden adinkra adjacency matrix element, $a^i_j$, picks up a minus sign if the respective edge is dashed, while $a^i_j =\alpha \frac{d}{d\tau}$ and $a^j_i =\alpha'$, if $\bullet_i \leftarrow \bullet_j$. Finally if an edge connects an open (boson) node to a closed (fermion) node the matrix element picks up a phase $\dot\imath$. For example, if bosons are labeled $\Phi_i$ and fermions, $\Psi_j$, the relations between nodes for Fig. \# 1(a) can be represented as\newline
\noindent
\begin{center}
$
~{\color{red}{\rm D}{}_{red}}
  \left( \begin{array}{c}
\Psi_1 \\ 
\Psi_2 \\ 
\end{array} \right)=\left( \begin{array}{cc}
0 & 1 \\ 
- 1&0\\ 
\end{array} \right) \left( \begin{array}{c}
\Phi_1 \\ 
\Phi_2 \\ 
\end{array} \right)\newline\newline\newline
~~~~~{\color{red}{\rm D}{}_{red}}
  \left( \begin{array}{c}
\Phi_1 \\ 
\Phi_2 \\ 
\end{array} \right)=\dot{\imath}\left( \begin{array}{cc}
0 & -\frac{d}{d\tau}\\ 
\frac{d}{d\tau}&0 \\ 
\end{array} \right) \left( \begin{array}{c}
\Psi_1 \\ 
\Psi_2 \\ 
\end{array} \right)
\newline\newline\newline
{\color{blue}{\rm D}{}_{blue}}
  \left( \begin{array}{c}
\Psi_1 \\ 
\Psi_2 \\ 
\end{array} \right)=\left( \begin{array}{cc}
1 & 0 \\ 
0& 1\\ 
\end{array} \right) \left( \begin{array}{c}
\Phi_1 \\ 
\Phi_2 \\ 
\end{array} \right)\newline
$
\end{center}
$
~~~~~~~~~~~~~~~~~~~~~{\color{blue}{\rm D}{}_{blue}}
  \left( \begin{array}{c}
\Phi_1 \\ 
\Phi_2 \\ 
\end{array} \right)=\dot{\imath}\left( \begin{array}{cc}
\frac{d}{d\tau}& 0\\ 
0 &\frac{d}{d\tau}\\ 
\end{array} \right) \left( \begin{array}{c}
\Psi_1 \\ 
\Psi_2 \\ 
\end{array} \right)~~~~~~~~~~~~~~~~~~~~~~~~~~~~~~~~~~(1)
$\newline
\newline\newline
\noindent
 and the equations for Fig. \# 1(b) can be represented as

\begin{center}
$%\newline
~~~{\color{red}{\rm D}{}_{red}}
  \left( \begin{array}{c}
\Psi_1 \\ 
\Psi_2 \\ 
\end{array} \right)=\left( \begin{array}{cc}
0&\frac{d}{d\tau} \\ 
-1 & 0\\ 
\end{array} \right) \left( \begin{array}{c}
\Phi_1 \\ 
\Phi_2 \\ 
\end{array} \right)\newline\newline\newline
~~~~{\color{red}{\rm D}{}_{red}}\left( \begin{array}{c}
\Phi_1 \\ 
\Phi_2 \\ 
\end{array} \right)=\dot{\imath}\left( \begin{array}{cc}
 0&-\frac{d}{d\tau}\\ 
1&0  \\ 
\end{array} \right) 
\left( \begin{array}{c}
\Psi_1 \\ 
\Psi_2 \\ 
\end{array} \right)\newline\newline\newline
{\color{blue}{\rm D}{}_{blue}}
  \left( \begin{array}{c}
\Psi_1 \\ 
\Psi_2 \\ 
\end{array} \right)=\left( \begin{array}{cc}
1 & 0 \\ 
0& \frac{d}{d\tau}\\ 
\end{array} \right) \left( \begin{array}{c}
\Phi_1 \\ 
\Phi_2 \\ 
\end{array} \right)\newline
$
\end{center}
$~~~~~~~~~~~~~~~~~~~~
~~{\color{blue}{\rm D}{}_{blue}}
  \left( \begin{array}{c}
\Phi_1 \\ 
\Phi_2 \\ 
\end{array} \right)=\dot{\imath}\left( \begin{array}{cc}
\frac{d}{d\tau}&0\\ 
 0& 1 \\ 
\end{array} \right) \left( \begin{array}{c}
\Psi_1 \\ 
\Psi_2 \\ 
\end{array} \right)~~~~~~~~~~~~~~~~~~~~~~~~~~~~~~~~~~~~~(2)$\newline\newline

\noindent
We can create a block matrix that makes the above notation more compact, called an adinkra adjacency matrix.
With the supervector ${\bf \Phi\oplus\Psi}$, the adinkra adjacency matrix for each edge color is of the form:\newline

$
~~~~~~~~~~~~~~~~~~~~~~~~~~~~~~~\left
\{\Gamma_I = 
\left( \begin{array}{cc}
{\bf 0} &{\bf D_{I{}_L}} \\ 
{\bf D_{I{}_R}} & {\bf 0}\\ 
\end{array} \right) \right\}
~\forall~ \text{\bf I} \in \{1,...,n+k\}~~~~~~~~~~~~~~~~~~~~~~~~~~~~~~~~(3)
$\newline\newline 
\noindent
where ${\bf D_{I{}_L}}$ permutes the the fermions, and ${\bf D_{I{}_R}}$, the bosons, in the above examples. The degrees of freedom of the generators can be symbolized by a subgraph called a baobab \cite{Baobab}. For example, the above graphs in Fig. \# 1 can be represented by their respective baobabs in Fig. \# 2.
 
 \begin{figure}[!htbp]
  \centering
   \subfigure[]{\label{f:bowtiemin}
   \includegraphics[width=0.3\columnwidth]{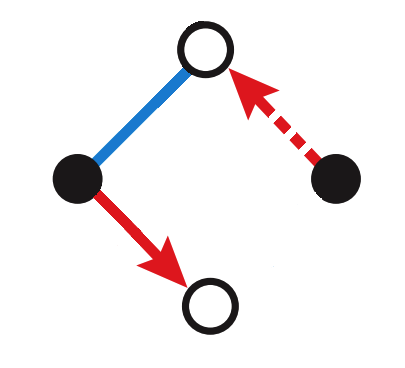}}
   \quad
   \subfigure[]{\label{f:diamondmin}
   \includegraphics[width=0.3\columnwidth]{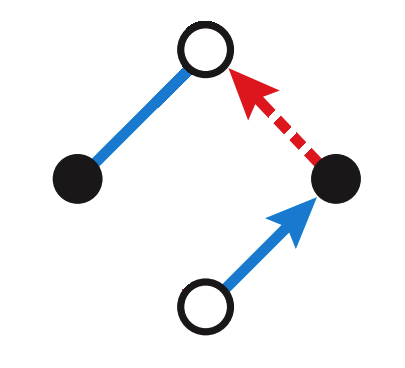}} 
   \label{f:diambtMin}
   \caption{An example of a Garden baobab for each adinkra in Fig. \# 1.}
\end{figure}

Lie baobabs so far explored seem to have relatively few cycles (this varies among superalgebras) implying that one can determine the weights of an adinkra in almost linear time compared to naively exploring every edge, which may take up to $\mathcal{O}(N \[N-1\])$ if the graph is $K_{N}$. 

Because Lie adinkras always contain some constrain with respect to their edge weights, the Lie baobab is always smaller than the adinkra. It is thus possible that many different Lie baobabs can create the same Lie adinkra. Determining the exact multiplicity of these graphs was an open question for all but the simplest examples. 

\section{The Spectrum Of Quotiented Hypercubes}

$~~\ ~$There are important exceptions, however. Ignoring edge directions, a Garden adinkra with $n$ edge colors and $2^n$ nodes contains:

$
~~~~~~~~~~~~~~~~~~~~~~~~~~~~~~~~~~~~~~~~~\sum\mathcal{B}(I^n) =\frac{1}{2^n}\prod_{j=1}^{n} (2~j)^{{n}\choose{j}}~~~~~~~~~~~~~~~~~~~~~~~~~~~~~~~~~~~~(4)\newline\newline
$
\noindent
spanning trees, and this an equal number of baobabs \cite{Combinatorics}, because those adinkras are $n$-cubes. This calculation was easily derived from the hypercube spectrum, therefore, one may intuitively expect that knowing the spectrum of adinkra topologies in general will help determine the multiplicity of the baobabs.
 Garden adinkras are either $n$-cubes or quotiented $n$-cubes \cite{Code}, therefore all we have to do is determine the spectrum of quotiented hypercubes. In this section, I will determine their spectrum, and explain some basic properties of this spectrum, before I use it to determine the number of spanning trees in the graphs.

 It has been well studied how a quotiented hypercube can be represented by a set of doubly even codewords, $\{C_i\}$. If each node is labeled on a unit hypercube by a boolean vector, $\mathbb{Z}^n_2$, then the quotient modulates the vectors: $\mathbb{Z}^n_2/C$ where $C$ is the set of doubly even codewords \footnote{See \cite{Code}, \cite{Code2}, \cite{Code3} or \cite{YanThesis} to understand the relation between doubly even codewords and Garden algebra.}. \newline\newline
{\bf Theorem 1}: The Laplace spectrum is: \newline\newline
$
~~~~~~~~~~~~~~~~~~~~~~~~~~~~~~~~~~~~\{\prod_{m=0}^n \prod_{p=0,~2(m+p)=const.}^k \[2(m+p)\]^{\mathcal{M} (m,p)}\}~~~~~~~~~~~~~~~~~~~~~~~~~~~~(5)
$\vskip0.2in
where $\lambda_i^N$ means that the eigenvalue $\lambda_i$ has degeneracy $N$. The regular spectrum is $\lambda_i-(n+k)$ because the graph is strongly regular \cite{Spectra}.
$\mathcal{M}(m,p)$ is the multiplicity for each $m$ and $p$, equal to:
\newline\newline
$
\mathcal{M} (m,p) =  \sum_{i\le k,~ p_i,~ p=\sum_j \[p_j (mod~2)\]}^{\|C_i\|-1} \sum_{q\le k,~ j\le q,~ i_{j-1} < i_j\le k-(q-j), ~p_{i_1...i_q}}^{p_{i_1...i_q} = \|C_{i_1}\wedge...\wedge C_{i_q}\|} \newline\newline
~~~~~~~~~~~\prod_{r=1}^k \prod_{j \le r,~i_j~=~1+i_{j-1}}^{i_j~=~k-(r-j)} {\|C_{i_1}\wedge...\wedge C_{i_r}\|  ~+ ~\sum_{q'\le k-r, j'\le q', \ell_{j'}\notin \{i_1,...,i_r\}}^{\ell_{j'} ~=~ k-(r-j')}  (-1)^{q'}~\|C_{i_1}\wedge...\wedge C_{i_r}\wedge...\wedge C_{\ell_{q'}}\| \choose p_{i_1...i_r} ~+~ \sum_{q'\le k-r, ~j'\le q', ~\ell_{j'}\notin \{i_1,...,i_r\}}^{\ell_{j'} = k-(r-j')}(-1)^{q'} ~p_{N\[i_1...i_r...\ell_{q'}\]}}\newline\newline
~~~~~~~~~~~~~~~~~~~~~~~~~~~~~~~~~~~~~~~~~~~~~~~{n+k + \sum_{q' \le k,~ j' \ell q, ~\ell_{j'} = 1+\ell_{j'-1}}^{\ell_{j'}=k-(q'-j')} (-1)^{q'} \|C_{i_1}\wedge...\wedge C_{i_{q'}}\|  \choose m+\sum_{q'\le k,~ j' \le q',~ \le_{j'}=1+\ell_{j'-1}}^{\ell_{j'}=k-(q'-j')}(-1)^{q'} ~p_{i_1...i_{q'}}}~~~~~~~~~~(6)
$\newline\newline
{\bf Proof:} See Appendix.\newline\newline
\noindent Here, $N[i_1 ... i_r ... \ell_{q'}]$ represents ordering lexicographically, (e.g. $p_{1 3 2} \rightarrow p_{1 2 3}$) and ``$\|C_i\|$" symbolized the weight of the code. As an aside, although the above formula may at first appear complicated, it can simplify into a few well known cases. For example, when $k=0$, the spectrum is that of a hypercube, as expected. In addition, when $N=4$, or when $N=8$, the quotiented hypercubes become complete bipartite graphs with 8 and 16 nodes, respectively. In both cases, one can compute by hand  that the eigenvalues are $\{0,~4^6\}$ and $\{0, ~8^{12},~16\}$. Lastly, Cartesian products of the above graphs are also known.

Thm. \# 1 implies an important property relating to the set of co-spectral graphs. Any codewords whose columns can be permuted from one to the other have the exact same spectrum, even when they represent completely distinct topologies. For example, the below codewords have their first and fifth columns switched and represent distinct topologies because they are not isomorphic. Their corresponding spectrums, however, are the same.
\newline\newline
$
~~~~~~~~~~\left[ \begin{array}{cccccccc}
 1&0&0&0&0&1&1&1 \\ 
1&0&1&1&0&1& 0&0\\ 
0&1&1&1&1&0&0&0
\end{array} \right]~~~~~~~~${\Huge$\leftrightarrow$}$~~~~~~~~~~~\left[ \begin{array}{cccccccc}
 0&0&0&0&1&1&1&1 \\ 
0&0&1&1&1&1& 0&0\\ 
1&1&1&1&0&0&0&0
\end{array} \right]~~~~~~~~(7)\newline\newline
$

 The spectrum is not, however, topologically invariant. As studied in \cite{Code}, the following set of codewords have distinct topologies {\it and} distinct spectra:\newline\newline
$
~~~~~~~~~~\left[ \begin{array}{cccccccc}
 0&0&0&0&1&1&1&1 \\ 
0&0&1&1&1&1& 0&0\\ 
0&1&0&1&0&1&0&1
\end{array} \right]~~~~~~~~${\Huge$\leftrightarrow$}$~~~~~~~~~~~\left[ \begin{array}{cccccccc}
 0&0&0&0&1&1&1&1 \\ 
0&0&1&1&1&1& 0&0\\ 
1&1&1&1&0&0&0&0
\end{array} \right]~~~~~~~~(8)\newline\newline
$
 \begin{figure}[!htbp]
  \centering
   \subfigure[]{\label{f:bowtiemin}
   \includegraphics[width=1\columnwidth]{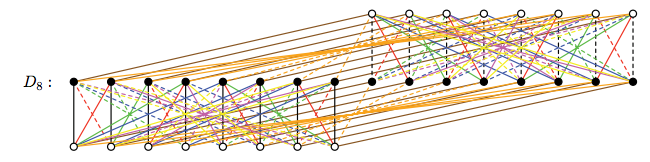}}
   \quad
   \subfigure[]{\label{f:diamondmin}
   \includegraphics[width=1\columnwidth]{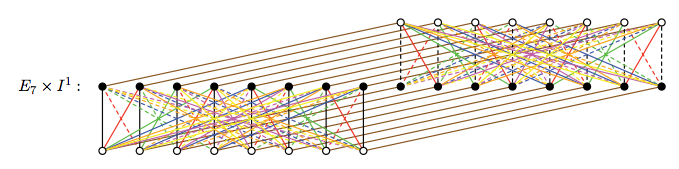}} 
   \label{f:diambtMin}
   \caption{The graph topologies corresponding to the respective codewords in (8).}
\end{figure}
\noindent
Trivially $\|C_i\|$, $\|C_i\wedge C_j\|$, ..., are distinct, and thus the spectra are distinct. Therefore a meta-equivalence class exists among codewords which are permutation {\it inequivalent} but non-trivially still have the same spectrum.

With these results, the overall appearance of the spectrum can be explored. Fig. \# 4 shows how the Laplace spectrum changes for an $n+k$-cube quotiented $k$ times. For $k=3$, 3 distinct codes were found, with a different width, but similar maxima. Interestingly the maximum multiplicity shifts to the right as $k$ increases. This agrees with ones intuition because $\lambda_{mode}\simeq \frac{n}{2}$ for a hypercube graph, and $2^{n-1}$ for a complete bipartite, thus as $k$ changes, there should be some sort of gradual shift in $\lambda_{mode}$. When $n+k$ is large, it can be shown that the change in $\lambda_{mode}$ is linear. Specifically,\newline\newline

\noindent
{\bf Proposition 1:}\newline\newline
$
~~~~~~~~~~~~~~~~~~~~~~~~~~~~~~~~~~~~~~~  \lambda_{mode} \simeq n+k,~n+k\gg ~1~~~~~~~~~~~~~~~~~~~~~~~~~~~~~~~~~~~~~(9)
$
\vskip0.2in
 \begin{figure}[!htbp]
  \centering
   \includegraphics[width=1\columnwidth]{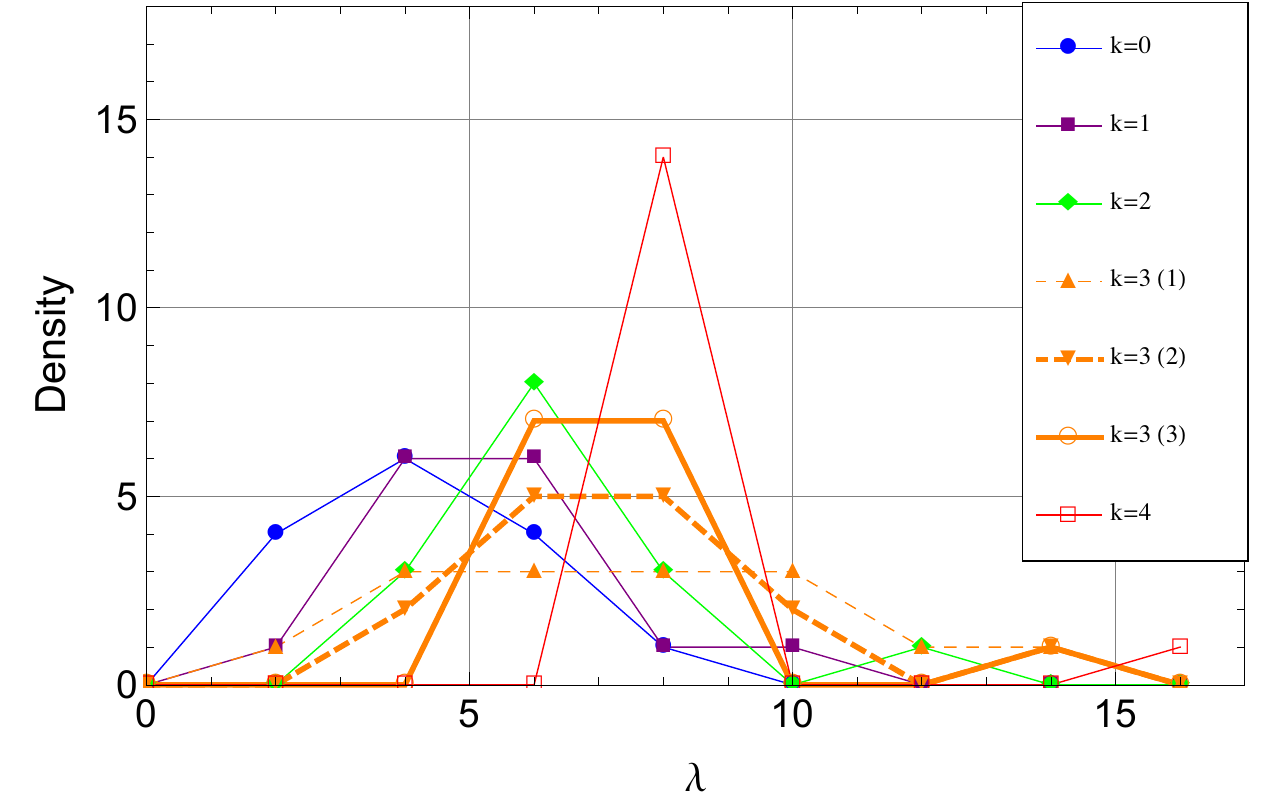}
   \caption{The Laplace spectrum for a $4+k$-cube quotiented $k$ times.}
\end{figure}
\noindent
\noindent
{\bf Proof:}\newline
To find the mode of the spectral density, \newline\newline
$
~~~~~~~~~~~~~~~~~~~~~~~~~~~~~~~~~~~~~~~  \partial_{\lambda} \sum_{p}  \mathcal{M}(\lambda/2-p,p) = 0~~~~~~~~~~~~~~~~~~~~~~~~~~~~~~~~~~~~~(10)
$
\vskip0.2in
  To remove the summation, the maximum multiplicity is explored\newline\newline
 $
 ~~~~~~~~~~~~~~~~~~~~~~~~~~~~~~~ \prod_{i_1} {\|C_{i_1}\|-1 - \sum_{i_{2'}}\|C_{i_1}\|\wedge \|C_{i_{2'}}\|+...\choose p_{i_1} - \sum_{i_{2'}} p_{i_1 i_{2'}}+...}\prod_{i_{2}}{\|C_{i_1}\wedge C_{i_{2}}\|-...\choose p_{i_1 i_{2}} - ...}...~~~~~~~~~~~~~~~~~~~~~~~~~~(11) $
 \vskip0.2in
 \noindent are simultaneously maximum when $p_{i_1}=(\|C_{i_1}\|-1)/2$, $p_{i_1 i_{2}} = \|C_{i_1} \wedge C_{i_{2}}\|/2$, etc. as can easily be proved by induction. Changing $p_1$, $p_2$,...$p_j$... by 1 makes a negligible change to the the binomial coefficient, although it significantly changes $p$. Given $p_1,...,p_k$, there are ${k\choose p}$ ways of changing $p_i$ such that $p$ is constant, hence $\mathcal{M}(\lambda/2-p,p)$ is maximal when $p=k/2$. Therefore, \newline
 \newline
 $ ~~\mathcal{M}_{peak}(\lambda/2-p,p) = {k\choose k/2}{\|C_{i_1}\| - \sum_{i_{2'}} \|C_{i_1}\wedge C_{i_2}\|+...\choose (\|C_{i_1}\| - \sum_{i_{2'}} \|C_{i_1}\wedge C_{i_{2'}}+...\|)/2}{\|C_{i_1}\wedge C_{i_{2}}\|-...\choose (\|C_{i_1}\wedge C_{i_{2}}\|-...)/2}... 
 \newline\newline~~~~~~~~~~~~~~~~~~~~~~~~~~~~~~~~~~~~~~~~~~~~~~~~~~~~~~~~~~~{n+k-\sum_{i_{1'}} \|C_{i_{1'}}\| + \sum_{i_{2'}>i_{1'}}\|C_{i_{1'}}\wedge C_{i_{2'}}\|-...\choose \lambda/2 -\sum_{i_{1'}} \|C_{i_{1'}}\|/2 + \sum_{i_{2'}>i_{1'}}\|C_{i_{1'}}\wedge C_{i_{2'}}\|/2-... }~~~~~(12)$
 \vskip0.2in
  Therefore, the partial derivative can be put inside the sum. By dropping values that are constant:\newline\newline
 $
~~~~~~~~~~~~~~~~~~~~~~~~~~~~~~~~~~~~~~~~\partial_\lambda {n+k-\alpha \choose \lambda/2 -\alpha/2} = 0\implies \lambda=n+k~~~~~~~~~~~~~~~~~~~~~~~~~~~~~~(13)
 $
 \vskip0.2in\noindent

 where $\alpha = \sum_{i_{1'}} \|C_{i_{1'}}\| - \sum_{i_{2'}>i_{1'}}\|C_{i_{1'}}\wedge C_{i_{2'}}\|+...$.
\begin{flushright}$\Box$\end{flushright}

Lastly, it should be noted that, by using the Gaborit mass formula \cite{MassFormula}, one can determine $k_{max}$ (the number of times in $N$-cube can be quotiented). This observation comes directly from research done by the DFGHIL collaboration \cite{Code}. For an $N$-cube, with $N=n+k = 8 m + s$, one can determine $k_{max}$ by the function:\vskip0.2in
$
~~~~~~~~~~~~ k_{max}(n+k) = \left\{ 
  \begin{array}{l l}
    4 m & \quad \text{for (n+k) $\equiv$ 0, 1, 2, or 3 $(mod ~8)$}\\
    4 m + 1 & \quad \text{for (n+k) $\equiv$ 4 or 5 $(mod ~8)$}\\
     4 m +2& \quad \text{for (n+k) $\equiv$ 6 $(mod ~8)$}\\
    4 m + 3 & \quad \text{for (n+k) $\equiv$ 7 $(mod~ 8)$}
      \end{array} \right.
  ~~~~~~~~~~~~~~~~(18)
$\vskip0.2in

\noindent Therefore, $C_{k>k_{max}}=00...$ implying $\mathcal{M}(m,p>k_{max})=0$ and $\mathcal{M}(m>n+k_{max},p)=0$.

The {\it Matrix-Tree Theorem} states that the number of spanning trees in a graph is $1/N \lambda_2...\lambda_N$. For a quotiented hypercube, this becomes:

$
~~~~~~~~~~~~~~~~~~~~~~~~~~~~~~~~ \sum\mathcal{T}(I^n/C) = \prod_m^n \prod_p^k \[ 2 (m+p)\]^{\mathcal{M}(m,p)} ~~~~~~~~~~~~~~~~~~~~~~~~~~~~~~~~~~~~~~~~(19)
$

 Because the baobab is made from spanning trees, bounds on the baobab multiplicity for a given adinkra can now be determined, ignoring edge directions.
 
 \section{The Entropy Of An Adinkra}
 
In this section the multiplicity of the baobabs (ignoring edge directions) will be shown to be \newline\newline
$
(\sum\mathcal{T}(I^n/C)) \frac{2^{(n-1) k}}{\lceil (2^n+k-1)/k \rceil^k}<\sum\mathcal{B}(I^n/C)
\newline\newline~~~~~~~~~~~~~~~~~~~~~~~~~~~~~~~\le (\sum\mathcal{T}(I^n/C)) \frac{2^{(n-1) k}}{\prod_{\forall q\le r,~i_1<...<i_q} \|C_{i_1} \wedge C_{i_2}\wedge...\wedge C_{i_q}\|^{(-1)^{(q-1)}}}~~~~~~~~~~(20)
$

This paper only focuses on the multiplicity of dashed edges and not directed edges in a baobab, due to the complications directed edges present. The multiplicity of directed edges is small for any valid baobab topology, but there are topologies, such as a Hamilton path in a hypercube graph, which are invalid under certain circumstances. The affect of these directed edges on the multiplicity is likely to be minimal, therefore they are safely ignored in this paper. 

As explain in \cite{Baobab}, a Garden baobab (which we will simply label a ``baobab" for the rest of the paper) is a spanning tree with $k$ cycles, each of which contains at least four colors that do not appear an even number of times. Furthermore, there exists an edge color in each cycle doesn't appear in any other cycle.  Because there are $2^{n-1}$ edges of the same color, there are $2^{n-1}$ equivalent cycles one can create a baobab from. Ignoring overcounting, this would imply the number of baobabs is $(\sum \mathcal{T}(I^n/C))2^{(n-1)k}$.

In truth, however, many spanning trees can create the same cycle (for example, a cycle $\mathcal{C}_N$ has $N$ spanning trees \cite{Spectra}). How many baobabs there really are is therefore an open question in general. Simple arguments, however, can set limits on the number of baobabs. If every edge was part of a cycle, then there could be no more than ${\  \lceil (2^n + k-1)/k\rceil}^k$ trees that make the same baobab. One can arrive at this number by imagining a baobab a collection of cycles $\{C_{N_1}, \mathcal{C}_{N_2},...,\mathcal{C}_{N_k}\}$ with exactly one node in common between any cycle and $N_1 + N_2 +... N_k= 2^n+k-1$. This value is clearly largest when $N_1=N_2=...=N_k$. Unrealistically, however, this assumes that the cycles always span the entire baobab. The number of trees in all baobabs is therefore less than $\lceil (2^n+k-1)/k \rceil^k$.

The minimum number of trees is when the $k$ cycles are smallest and have the most edges in common. This happens when all cycles are of size $\|C_{i_1}\|$ (where $\|C_{i_1}\|$ is the weight of the codeword), and have $\|  C_{i_1}\wedge C_{i_2}\wedge...\|$ edges in common with one another. The total number of trees is therefore less than or equal to $\prod_{\forall q\le r,~i_1<i_2<...i_q} \|C_{i_1} \wedge C_{i_2}\wedge...\wedge C_{i_q}\|^{(-1)^{(q-1)}}$. This implies (18).

It is given that when $k=0$, the upper and lower bounds are the same. In addition, one can see that when $n=3$ and $k=1$, the total number of baobabs is equal to $(\sum\mathcal{T}(I^3_{~1})) \frac{2^{2}}{\|C_{1}\|}$ implying the upper bounds is tight, although simple constructions show that the bound is not tight in general. It seems likely that the value approaches the lower bounds in the limit of $k\ll n$ while it approaches the upper bounds when $k\sim n$.

The log of the baobab multiplicity is the Shannon entropy of the adinkra. Once the adinkra is fully constructed, information is lost about the original baobab. The second law of thermodynamics can be used to determine the minimum energy needed to create an adinkra \cite{Baobab}:\vskip0.2in
$~~~~~~~~~~~~~~~~~~~~~~~~~~~~~~~~~~~~~~~~~~~\delta Q = \Delta S ~T~~~~~~~~~~~~~~~~~~~~~~~~~~~~~~~~~~~~~~~~~~~~~(26)$
\vskip0.2in This value can be regarded as the latent heat needed for a phase transition from the baobab to the adinkra. Although this was derived for Garden algebra, the multiplicity and hence the entropy and heat generation is seen for all Lie adinkras.

 \section{Thermodynamics Of Adinkras}

 By interpreting baobabs as ``microstates", and the complete adinkra as the ``macrostate" the multiplicity of an adinkra can be defined as as the number of baobabs that can create same adinkra. This definition is equivalent to Shannon entropy by noting that information is lost under the surjective transformation from a baobab to the adinkra. Just as logic gate turns 2 bit of information into 1 bit, and therefore looses information about the initial input, an adinkra cannot reconstruct the initial baobab and thus looses information. 

Note that the logic gate concept of adinkras only works by assuming the topology of the baobab is known.
In other words, to create the logic gate set-up from the baobab, organize all edges as ordered pairs of nodes with a bit to determine dashing $\mathbb{Z}^n_2/C \times \mathbb{Z}^n_2/C \times \{0,1\} = (u,v,x)$. The logic gates for the dashing of a baobab are all functions NDXOR such that:
\newline\newline $NDXOR_{baobab}:\newline\newline~~~~~~~~~~~~~~~\{(u,u\oplus C_i,x), (u,v,y), (v,v\oplus C_i,z)\}~~${\Huge$\rightarrow$}~~~$(u\oplus C_i,v\oplus C_i,\neg (x\oplus y\oplus z))~~~~~(27)$\newline\newline
 
Once the logic-gate setup is known, edge dashing can be fed into the gates as input bits. This implies that the Shannon entropy  is due to the uncertainty of the baobab's construction.

Because $\sum \mathcal{T}(I^n/C)> \sum \mathcal{T}(I^n)>2^{2^n}\gg 2^{(n-1) k}>\frac{\sum \mathcal{B}(I^n/C)}{\sum \mathcal{T}(I^n/C)}$ for large $n$ and $k$:
\newline\newline
 $~~~~~~~~~~~~~~~~~~~~~~~~~~~~~~~~S = k_b~ \text{ln}(\sum \mathcal{B})\simeq k_b~ \text{ln} (\sum T (I^n/C))+\mathcal{O}(k~ \text{ln}(k))~~~~~~~~~~~~~~~~~~~~~~~~~~~~~(26)$ 
 \vskip0.2in\noindent
 to a good approximation, if we assume it is equal probable for any valid baobab to create a given adinkra.

Because the entropy of an adinkra is known, so is the minimal energy necessary to construct an adinkra from a baobab. Recall for small changes in energy for systems that are approximately isothermal, the heat, 
\vskip0.2in
$~~~~~~~~~~~~~~~~~~~~~~~~~~~~~~~~~~~~~~~~~~~\delta Q \ge k_b~\text{ln}(\sum \mathcal{T}(I^n/C)) ~T~~~~~~~~~~~~~~~~~~~~~~~~~~~~~~~~~~~~~~~~~~~~~(27)$
\vskip0.2in
\noindent where $\Delta S$ is the small change in entropy, and $T$ is temperature. Therefore, we can find the minimal heat lost in order to create the adinkra by the value $\Delta S ~ T$.

\section{Conclusion}

In this paper, I introduced the spectrum of a quotiented hypercube, which to the knowledge of the author, has never appeared previously in literature. Furthermore, using the Laplace spectrum, the number of spanning trees was found, allowing for a strong bound on the number of baobabs in an adinkra, if directed edges are ignored. Lastly, knowing the multiplicity of the baobab, one can determine the Shannon entropy, as well as the minimal energy necessary to create an adinkra. This allows for the interpretation that, by completing 2-color loops, one can create a phase transition between a baobab and an adinkra with some latent heat.

There is much left to study, however. Edge directions were completely ignored, for example. By taking edge directions into account, bounds on the multiplicity of baobabs can be significantly improved. Furthermore, it is an open question whether a similar argument can determine the spectrum and thermodynamics of other Lie adinkras.

\section{Appendix}

Equations (5) and (6) will be proven using a construction provided below.
\vskip0.2in
{\bf Construction 1:} Any $(n,k)$ set of codewords can be put into a form where $\forall~ i,~\|\alpha'_i\| = \|C_i - \sum_{j\neq i} C_i\wedge C_j + \sum_{\ell>j,~\neq i} C_i\wedge C_j\wedge C_\ell-... \|>0$. This means each code will contain at least one column where, $\forall ~i$, the $i^{th}$ row contains the only bit.
\begin{itemize}

\item If $\|\alpha'_i\|=0$, then choose some bit $x\in C_i$. $\forall~ i\neq j$ where $\|0...0x0...0\wedge C_j\|\neq 0$, $C_j\rightarrow C_j\oplus C_i$. It is easy to see that this forces $\|\alpha'_i\|>0$.

\item If any other rows contain $\|\alpha'_j\|=0$, then repeat the first step (by the construction $\|\alpha'_i\|$ will stay positive definite).

\end{itemize}

This construction allows us to create a nice form from which to determine the multiplicity of baobabs (ignoring directed edges). It is easy to see from the act of quotienting that an edge, $\Gamma_l = \Gamma_a \Gamma_b \Gamma_c...$ where $\{\Gamma_l,~\Gamma_a,~\Gamma_b,~\Gamma_c,...\} \in C_i$. Therefore, while hypercube edges are usually of weight 1, quotiented hypercubes are of weight 1 or $\| C_i\|-1$. By choosing $\Gamma_a\Gamma_b \Gamma_c... \in C_i-x$, where $x\in \alpha'$, $u\wedge C_i \equiv u\wedge \Gamma_l~(mod~2)$ (without this construction, $\Gamma_a$ could be equal to $\Gamma_e \Gamma_f \Gamma_g...$ where $\{\Gamma_a,~\Gamma_e,~\Gamma_f,~\Gamma_g,...\} \in C_j$, implying $u\wedge C_i$ may or may not equal $u\wedge \Gamma_l$).
\newline\newline
\vskip0.2in
{\bf Theorem 1:} The Laplace spectrum of a quotiented hypercube is (5) with multiplicity (6) and $k\le k_{max}$.\newline\newline
{\bf Proof:}\newline\newline

All nodes can be labeled as a quotients of $n+k$ boolean vectors $\mathbb{Z}^n_2/C$. All edges, by construction \# 1, can be codewords of weight 1 or $\|C_i\|-1$, where the weight $\|C_i\|-1$ codewords are equal to $C_i\oplus y$ where $y\in \alpha'_i$. 

Fix two vertices $u$ and $v$, and let $p_i=\|u\wedge (C_i\oplus y)\|$. When $p_i=a (mod~ 2)$, $(-1)^{u\cdot v} = a (-1)^{u \cdot w}$. Here, $\cdot$ is the inner product of the boolean vectors (i.e. $\|u\wedge v \|$), $w\sim v$ and $a=\{0,1\}$. $\|u\| = p_1 + p_2 +...+p_k-\beta$, where $\beta$ is some number that will be determined later, and $p$ variables will be $1 (mod~ 2)$ while $k-p$, $0 (mod~ 2)$. 
 Because $k$ weight 1 codewords became weight $\|C_i\|-1$ codewords, $n$ of $v$'s neighbors have a weight that differs by 1. $(-1)^{u \cdot v} = -(-1)^{u \cdot w}$ for $m=\|u\|$ neighbors of $v$, and $(-1)^{u \cdot v} = (-1)^{u \cdot w}$ for $n-m$ neighbors of $v$. Similarly, from the previous observation, $(-1)^{u \cdot v} = -(-1)^{u \cdot w}$ for $p$ neighbors of $v$, and $(-1)^{u \cdot v} = (-1)^{u \cdot w}$ for $k-p$ neighbors of $v$. As an ansatz, let $f_u = \{(-1)^{u\cdot v}\}$ be the eigenvector. Therefore, the $v^{th}$ row of $L (f_u)$ is \newline\newline
$
~~~~~~~~~~~~~~~~~~~~~~~~L (f_u)\bigg{|}_v = (n+k) (-1)^{u \cdot v} - \sum_{<v, w>} (-1)^{u \cdot w} ~~~~~~~~~~~~~~~~~~~~~~~~~~~~(21)
\newline\newline~~~~~~~~~~~~~~~~~= (n+k) (-1)^{u \cdot v} - (n-m-k-p) (-1)^{u\cdot v} + (k+p) (-1)^{u \cdot v}
~~~~~~~~~~~~~~~~~~~(22)
$\vskip0.2in
\noindent
Simplifying, $L (f_u) = 2 (m+p) f_u$. By construction the eigenvectors are clearly linearly independent, therefore our ansatz is valid.
 The multiplicity is slightly more complicated to compute. The multiplicity of the portion of $u \in \alpha'_i$ is simply: \newline\newline
$
~~~~~~~~~~~~~~~~~~~~~~~~~~~~~~~~~~~~{\alpha'_{i_1} \choose p_{i_1} - \sum_{i_{2'}\neq i_1} p_{i_1 i_{2'}} + \sum_{i_{3'}>i_{2'}, i_3\neq i_1} p_{i_1 i_{2'} i_{3'}}-...}~~~~~~~~~~~~~~~~~~~~~~~~~~~~~~~~~~~~~~(23)
$\vskip0.2in\noindent
where $p_{i_1 i_{2'}}= u \cdot (C_{i_1}\wedge C_{i_{2'}}) \le \|C_{i_1}\wedge C_{i_{2'}}\|$, and $p_{i_1 i_{2'} i_{3'}}= u \cdot (C_{i_1}\wedge C_{i_{2'}}\wedge C_{i_{3'}}) \le \|C_{i_1}\wedge C_{i_{2'}}\wedge C_{i_{3'}}\|$,.... The sum of $ p_{i_1} - \sum_{i_{2'}\neq i_1} p_{i_1 i_{2'}} + \sum_{i_{3'}>i_{2'}, i_3\neq i_1} p_{i_1 i_{2'} i_{3'}}-...$ in the binomial coefficient is simply $u \cdot \alpha'_i$. By summing $p_{i_1 i_2 ...} = 0,~1,~2,...$ such that $p = \sum_j \[ p_j (mod~ 2)\]$ one can find the total multiplicity. Similarly, the multiplicity of the bits affected by $C_{i_1}\wedge C_{i_2}$ alone is: \newline\newline
$
~~~~~~~~~~~~~~~~~~~~~~~~~~~~{\|C_{i_1}\wedge C_{i_{2}}\| - \sum_{i_{3'}\neq \{i_1i_2\}} \|C_{i_1}\wedge C_{i_{2}}\wedge C_{i_{3'}} \| +...\choose p_{i_1i_2} - \sum_{i_{3'}\neq \{i_1i_2\}}p_{i_1i_2i_{3'}}-...}~~~~~~~~~~~~~~~~~~~~~~~~~~~~~~~~(24) 
$\vskip0.2in\noindent
which is independent of (23). Continuing in this fashion, one can determine the multiplicity up until $C_1\wedge C_2\wedge...\wedge C_k$. The total multiplicity of the bits affected by quotienting is the product of all these multiplicities. To determine the multiplicity of the bits unaffected by the quotienting, the simple relation below is used:
\newline\newline
$
~~~~~~~~~~~~~~~~~~~~~~~~{n+k -\alpha \choose m-\sum_{i_{1'}} p_{i_{1'}} +\sum_{i_{2'} \neq i_{1'}} p_{i_{1'} i_{2'}} -...}~~~~~~~~~~~~~~~~~~~~~~~~~(25)
$
 Therefore the spectrum is (5) and (6).

\begin{flushright}$\Box$\end{flushright}
 
{\Large\bf Acknowledgments}

KB would like to thank Dr. S. James Gates of the University of Maryland for his many stimulating discussions. Adinkras and baobabs were drawn with the help of {\it Adinkramat} $\copyright$ 2008 by G. Landweber.

\end{document}